\documentclass[12pt]{JHEP3}
\usepackage{amsmath,epsfig}
\def\be{\begin{equation}}
\def\ee{\end{equation}}
\def\ba{\begin{eqnarray}}
\def\ea{\end{eqnarray}}

\def\beq{\begin{equation}}
\def\eeq{\end{equation}}
\def\beqa{\begin{eqnarray}}
\def\eeqa{\end{eqnarray}}

\def\yzero{\smash{\hbox{$y\kern-4pt\raise1pt\hbox{${}^\circ$}$}}}

\def\beq{\begin{equation}}
\def\eeq{\end{equation}}
\def\beqa{\begin{eqnarray}}
\def\eeqa{\end{eqnarray}}

\def\-{\hphantom{-}}

\def\s2{\frac{1}{\sqrt2}}

\def\beq{\begin{equation}}
\def\eeq{\end{equation}}
\def\beqa{\begin{eqnarray}}
\def\eeqa{\end{eqnarray}}

\def\IF{\relax{\rm I\kern-.18em F}}
\def\II{\relax{\rm I\kern-.18em I}}
\def\IP{\relax{\rm I\kern-.18em P}}
\def\IC{\relax\hbox{\kern.25em$\inbar\kern-.3em{\rm C}$}}
\def\IR{\relax{\rm I\kern-.18em R}}

\def\Dsl{\,\raise.15ex\hbox{/}\mkern-13.5mu D} 
\def\IZ{Z\kern-.4em  Z}

\title{N=1, 4D Supermembrane from 11D.}

\author{M. P. Garc\'\i a del Moral $^1$, J. M. Pe\~na $^2$, A. Restuccia $^3$\,\footnote{E-mail:
\emph{garcia@to.infn.it;
jmpena@ciens.fisica.ucv.ve; restucci@aei.mpg.de,arestu@usb.ve}}\\
$^1$ Dipartimento di Fisica Teorica, Universit\`a di Torino \\
and INFN - Sezione di Torino; Via P. Giuria 1; I-10125 Torino,
Italy.\\ \\
$^2$ Departamento de F\'\i sica, Facultad de Ciencias,\\
 Universidad Central de Venezuela,
 A.P. 47270, Caracas 1041-A, Venezuela\\ \\
$^3$ Max-Planck-Institut f\"ur Gravitationphysik,\\ Albert-Einstein-Institut
M\"ulenberg 1, D-14476 Potsdam, Germany\\ $\&$
Departamento de F\'\i sica, Universidad Sim\'on Bol\'\i var\\
Apartado 89000, Caracas 1080-A, Venezuela}

\abstract{ The action of the 11D supermembrane with nontrivial
central charges compactified on a 7D toroidal manifold is obtained.
It describes a supermembrane evolving in a 4d Minkowski space-time.
The action is invariant under additional symmetries in comparison to
the supermembrane on a Minkowski target space. The hamiltonian in
the LCG is invariant under conformal transformations on the Riemann
surface base manifold.  The spectrum of the regularized hamiltonian
is discrete with finite multiplicity. Its resolvent is compact. Susy
is spontaneously broken, due to the topological central charge
condition, to four supersymmetries in 4D, the vacuum belongs to an
N=1 supermultiplet. When assuming the target-space to be an
isotropic 7-tori, the potential does not contain any flat direction,
it is stable on the moduli space of parameters.}

\preprint{DFTT-21/2007\\
AEI-2007-142}
\keywords{supermembrane, 4D,
moduli, supersymmetry}

\begin{document}
\section{Introduction}
The non-perturbative quantization of String Theory is still an open
problem which has received recently a lot of attention. It can be
reformulated in terms of the quantization of the M-theory in 11
dimensions. In spite of the advances towards its quantization
(\cite{dwhn}-\cite{bgmr2}), M-theory is a not well understood
theory. The goal to achieve is to obtain a consistent quantum theory
in 4D with an $N=1$ or $N=0$ supersymmetries, moduli free, in
agreement with the observed 4D physics. Attempts to formulate the
theory in 4D have been done in the supergravity approach
\cite{gutowski, lukas}, including fluxes \cite{acharya, dall'agata},
but so far no exact formulation has been found. The theory when
compactified to a 4D model contains many vacua due to the presence
of moduli fields. Stabilizacion of these moduli  is an important
issue to be achieved. For some interesting proposals see
\cite{acharya, diana}. In the following we will consider the $11D$
M2-brane with irreducible wrapping on the compact sector of the
target manifold \cite{ovalle}. This implies the existence of a non
trivial central charge in the supersymmetric algebra. Its spectral
properties have been obtained in several papers
\cite{gmr}-\cite{bgmr2}. Summarizing them: classically it does not
contain singular configuration with zero energy, and quantically its
regularized bosonic and supersymmetric hamiltonians have a discrete
spectrum with finite multiplicity. This proof has been extended to
the exact case for the bosonic sector of a supermembrane with
central charges compactified in a 2-torus. Extensions of this proof
to the supersymmetric case and other target manifolds are currently
under study.

The purpose of this paper is to construct the action for the
supermembrane with nontrivial central charges compactified on a
$T^{7}$ and analyze its physical properties. The action describes a
supermembrane evolving in a 4D Minkowski space. It is invariant
under SUSY with a Majorana 32 component spinor parameter however the
symmetry is spontaneously broken to a $N=1$ theory in 4  dimensions
when the minimal
configuration is fixed. A detailed analysis of the compactification
properties will be given elsewhere \cite{belhaj}. When the compactification
manifold is considered to
be an isotropic $T^{7}$ i.e. all the radii are equal,  the potential has no
flat directions, it is
stable in the moduli space of parameters.
\newline

The paper is structured as follows:  section 2 is devoted to explain
the construction of the supermembrane with central charges. In
section 3 we show how the compactification on the remaining
7-coordinate is performed to obtain a $4D$ action of the
supermembrane. Section 4 shows explicitly the spontaneous breaking
of supersymmetry. In section 5 we perform the analysis of the
spectral properties of the theory. In section 6 we analyze some
physical properties as mass generation and moduli stabilization due
to the topological condition and finally we present our conclusions
in section 7.

\section{D=11 Supermembrane with  central charges on a $M_{5}\times T^{6}$ target manifold}

The hamiltonian of the $D=11$ Supermembrane \cite{bst} may be
defined in terms of maps $X^{M}$, $M=0,\dots, 10$, from a base
manifold $R\times \Sigma$, where $\Sigma$ is a Riemann surface of
genus $g$ onto a target manifold which we will assume to be $11-l$
Minkowski $\times$ $l$-dim Torus. The canonical reduced hamiltonian
to the light-cone gauge has the expression

\begin{equation}\label{e1}
   \int_\Sigma  \sqrt{W} \left(\frac{1}{2}
\left(\frac{P_M}{\sqrt{W}}\right)^2 +\frac{1}{4} \{X^M,X^N\}^2+
{\small\mathrm{\ Fermionic\ terms\ }}\right)
\end{equation}
subject to the constraints \begin{equation}  \label{e2}
\phi_{1}:=d(\frac{p_{M}}{\sqrt{W}}dX^{M})=0 \end{equation} and
\begin{equation} \label{e3}
\phi_{2}:=
   \oint_{C_{s}}\frac{P_M}{\sqrt{W}} \quad dX^M = 0,
\end{equation}
where the range of $M$ is now $M=1,\dots,9$ corresponding to the
transverse coordinates in the light-cone gauge, $C_{s}$,
$s=1,\dots,2g$ is a basis of  1-dimensional homology on $\Sigma$,

 \be \label{e4}\{X^{M}, X^{N}\}= \frac{\epsilon
^{ab}}{\sqrt{W(\sigma)}}\partial_{a}X^{M}\partial_{b}X^{N}. \ee

$a,b=1,2$ and $\sigma^{a}$ are local coordinates over $\Sigma$.
$W(\sigma)$
 is a scalar density introduced in the light-cone gauge fixing procedure.
 $\phi_{1}$ and $\phi_{2}$ are generators of area preserving diffeomorphisms. That is
\be \sigma\to\sigma^{'}\quad\to\quad W^{'}(\sigma)=
W(\sigma).\nonumber \ee When the target manifold is simply connected
$dX^{M}$ are exact one-forms.

The $SU(N)$ regularized model obtained from (\ref{e1}) was shown to
have conti\-nuous spectrum from $[0,\infty)$,
\cite{dwln},\cite{dwmn},\cite{dwhn}.

This property of the theory relies on two basic facts: supersymmetry
and the presence of classical singular configurations, string-like
spikes, which may appear or disappear without changing the energy of
the model but may change the topology of the world-volume. Under
compactification of the target manifold generically the same basic
properties are also present and consequently the spectrum should be
also continuous \cite{dwpp}. In what follows we will impose a
topological restriction on the configuration space. It characterizes
a $D=11$ supermembrane with non-trivial central charges generated by
the wrapping on the compact sector of the target space
\cite{gmr},\cite{bgmmr},\cite{bgmr},\cite{bgmr2}. Following
\cite{bellorin} we may extend the original construction on a
$M_{9}\times T^{2}$ to $M_{7}\times T^{4}$, $M_{5}\times T^{6}$
target manifolds by considering genus $1,2,3$ Riemann surfaces on
the base respectively. Under such correspondence there exists a
minimal holomorphic immersion from the base to the target manifold.
The image of $\Sigma$ under that map is a calibrated submanifold of
$T^{2}, T^{4}, T^{6}$ respectively. The model in those cases present
additional interesting symmetries beyond the original ones
\cite{bgmr2}.

We are interested in reducing the theory to a 4 dimensional model,
we will then assume a target manifold $M_{4}\times T^{6}\times
S^{1}$. The configuration maps satisfy:

\begin{equation}\label{e5}
 \oint_{c_{s}}dX^{r}=2\pi S_{s}^{r}R^{r}\quad r,s=1,\dots,6.
\end{equation}\begin{equation}\label{e6} \oint_{c_{s}}dX^{7}=2\pi L_{s}R
\end{equation} \begin{equation}\label{e7}
 \oint_{c_{s}}dX^{m}=0 \quad m=8,9
\end{equation}
\\

 where $S^{r}_{s}, L_{s}$ are integers and $R^{r}, r=1,\dots,6$ are the radius of $T^{6}=S^{1}\times\dots\times S^{1}$
while $R$ is the radius of the remaining $S^{1}$ on the target. This
conditions ensure that we are mapping $\Sigma$ onto a
$\Pi_{i=1}^{7}S_{i}^{1}$ sector of the target manifold.

We now impose the central charge condition \be\label{e8}
I^{rs}\equiv \int_{\Sigma}dX^{r}\wedge dX^{s}=(2\pi
R^{r}R^{s})\omega^{rs} \ee where $\omega^{rs}$ is a symplectic
matrix on the $T^{6}$ sector of the target.

For simplicity we take $\omega^{rs}$ to be the canonical symplectic
matrix
\\

\begin{equation}
     M=\begin{pmatrix} 0 & 1 & & & &\\
                       -1 & 0 & & & & \\
                       & & 0 & 1 & & \\
                       & & -1& 0 & & \\
                       & & & & 0 & 1 \\
                       & & & & -1 & 0
 \end{pmatrix}.
\end{equation}
It corresponds to the orthogonal intersection of the three
$g=1$,i.e. three toroidal
 $T^{2}$ supermembranes, the time direction being the intersecting space.
 The topological condition
(\ref{e8}) does not change the field equations of the hamiltonian
(\ref{e1}).
  In fact, any variation of $I^{rs}$ under a change $\delta X^{r}$, single valued over
 $\Sigma$, is identically zero. In addition to the field equations obtained from (\ref{e1}),
the classical configurations must satisfy the condition (\ref{e8}).
It is only a topological restriction on the original set of
classical solutions of the field equations. In the quantum theory
the space of physical configurations is also restricted by the
condition (\ref{e8}). The geometrical interpretation of this
condition has been discussed in previous work
\cite{torrealba},\cite{ovalle}. We noticed that (\ref{e8}) only
restricts the values of
 $S_{s}^{r}$, which are already integral numbers from (\ref{e5}).\\

We consider now the most general map satisfying condition
(\ref{e8}). A closed one-forms $dX^{r}$ may be decomposed into the
harmonic plus exact parts: \begin{equation}
dX^{r}=M_{s}^{r}d\widehat{X}^{s}+dA^{r}
\end{equation}where $d\widehat{X}^{s}$, $s=1,\dots,2g$ is a basis of
harmonic one-forms over $\Sigma$. We may normalize it by choosing a
canonical basis of homology and imposing

\begin{equation} \oint_{c_{s}}d\widehat{X}^{r}=\delta_{s}^{r}. \end{equation} We have now
considered a Riemann surface with a class of equivalent canonical
basis. Condition (\ref{e5}) determines \begin{equation}\label{eq1}
M_{s}^{r}=2\pi R^{r}L_{s}^{r}.\end{equation} $dA^{r}$ are exact
one-forms. We now impose the condition (\ref{e8}) and obtain
\begin{equation} S_{t}^{r}\omega^{tu}S_{u}^{s}=\omega^{rs}, \end{equation} that
is, $S\in
Sp(2g,Z)$. This is the most general map satisfying (\ref{e8}).\\

 The
natural election for $\sqrt{W(\sigma)}$ in this geometrical setting
is to consider it as the density obtained from the pull-back of the
Kh\"aler two-form on $T^{6}$. We then define \begin{equation}
\sqrt{W(\sigma)}=\frac{1}{2}\partial_{a}\widehat{X}^{r}\partial_{b}\widehat{X}^{s}\omega_{rs}.
\end{equation}

$\sqrt{W(\sigma)}$ is then invariant under the change
\begin{equation} d\widehat{X}^{r}\to S_{s}^{r}d\widehat{X}^{s}, \quad
S\in Sp(2g,Z) \end{equation}

But this is just the change on the canonical basis of harmonics
one-forms when a biholomorphic map in $\Sigma$ is performed changing
the canonical basis of homology. That is, the biholomorphic (and
hence diffeomorphic) map associated to the mo\-du\-lar
transformation on a Teichm\"uller space. We thus conclude that the
theory  is invariant not only under the diffeomorphisms generated by
$\phi_{1}$ and $\phi_{2}$ but also under the diffeomorphisms,
biholomorphic maps,
changing the canonical basis of homology by a modular transformation.\\

The theory of supermembranes with central charges in the light cone
gauge (LCG) we have constructed depends then on the moduli space of
compact Riemanian surfaces $M_{g}$ only. It may be defined on the
conformal equivalent classes of compact Riemann surfaces. It shares
this property with String Theory, although the supermembrane theory
is still restricted by the area preserving constraints, there are
area preserving diffeomorphisms which are not conformal mappings. In
addition, the supermembrane depends on the moduli identifying the
holomorphic immersion from $M_{g}$ to the target manifold. This is
an interesting moduli space already considered in a different
context in \cite{witten}. \\

Having identified the modular invariance of the theory we may go
back to the general expression of $dX^{r}$, we may always consider a
canonical basis in away that \begin{equation}
dX^{r}=R^{r}d\widehat{X^{r}}+dA^{r}. \end{equation} the
corresponding degrees of freedom are described exactly by the
single-valued
 fields $A^{r}$. After replacing this expression in the hamiltonian (\ref{e1}) we obtain,

\begin{equation}\label{e9}
 \begin{aligned}
H&=\int_{\Sigma}\sqrt{W}d\sigma^{1}\wedge
d\sigma^{2}[\frac{1}{2}(\frac{P_{m}}{\sqrt{W}})^{2}+\frac{1}{2}
(\frac{\Pi^{r}}{\sqrt{W}})^{2}+
\frac{1}{4}\{X^{m},X^{n}\}^{2}+\frac{1}{2}(\mathcal{D}_{r}X^{m})^{2}+\frac{1}{4}(\mathcal{F}_{rs})^{2}
\\&+ \int_{\Sigma} \mathcal{F}+(2\pi)^{4}(R^{r}R^{s})^{4}\int_{\Sigma}\sqrt{W}(D_{r}\widehat{X})^{2}+\Lambda
(\mathcal{D}_{r}(\frac{\Pi_{r}}{\sqrt{W}})+\{X^{m},\frac{P_{m}}{\sqrt{W}}\})]\\& + \int_{\Sigma}
\sqrt{W} [- \overline{\Psi}\Gamma_{-} \Gamma_{r}
\mathcal{D}_{r}\Psi] +\overline{\Psi}\Gamma_{-}
\Gamma_{m}\{X^{m},\Psi\}+
 \Lambda \{ \overline{\Psi}\Gamma_{-},\Psi\}]
 \end{aligned}
 \end{equation}
where  $\mathcal{D}_r X^{m}=D_{r}X^{m} +\{A_{r},X^{m}\}$,
$\mathcal{F}_{rs}=D_{r}A_s-D_{s }A_r+ \{A_r,A_s\}$, \\
 $D_{r}=2\pi
R^{r}\frac{\epsilon^{ab}}{\sqrt{W}}\partial_{a}\widehat{X}^{r}\partial_{b}$
and $P_{m}$ and $\Pi_{r}$ are the conjugate momenta to $X^{m}$ and
$A_{r}$ respectively. $\mathcal{D}_{r}$ and $\mathcal{F}_{rs}$ are
the covariant derivative and curvature of a symplectic
noncommutative theory \cite{ovalle},\cite{bgmmr}, constructed from
the symplectic structure $\frac{\epsilon^{ab}}{\sqrt{W}}$ introduced
by the central charge. We  will take the integral of the curvature
to be zero and the volume term corresponds to the value of the
hamiltonian at its ground state. The last term represents its
supersymmetric extension in terms of Majorana spinors. The physical
degrees of the theory are the $X^{m}, A_{r}, \Psi_{\alpha}$ they are
single valued fields on $\Sigma$.

In \cite{bellorin} a $T^{4}$ compactification sector was considered.
Its hamiltonian was expressed in terms of a different frame for the
compactified  sector on the $T^{4}$ torus. In that case the pullback
is performed directly with the harmonic modes $d\widehat{X}^{r}$
while in the present formulation the metric on that sector is
$\delta_{rs}$ and the pullback should be performed with
$G^{1/2}d\widehat{X}$, $G$ is the constant matrix introduced in
\cite{bellorin}. In both cases the same scalar density
$\sqrt{W(\sigma)}$ is obtained.

\section{Compactification on the remaining $S^{1}$}

The analysis of the compactification on the remaining $S^{1}$ may be
performed directly in the
 above formalism or by considering its dual formulation in term of $U(1)$ gauge fields.
 We will discuss both approaches.

 In the first case, we may solve the condition (\ref{e6}), we obtain
 \begin{equation}\label{e10}
 dX^{7}= R L_{s}d\widehat{X}^{s}+d\widehat{\phi}
 \end{equation}
 where $d\widehat{\phi}$ is an exact 1-form and $d\widehat{X}^{s}$
 as before are a basis of harmonic 1-forms over $\Sigma$.
 For the discretness analysis at is more convenient to express $dX^{7}$
 in terms of the solution of the ``covariant'' laplacian over $\Sigma$:


 \begin{equation}\label{e11}
 \mathcal{D}_{r}\mathcal{D}_{r}\widetilde{X}=0
\end{equation}
where $\mathcal{D}_{r}$, $r=1,\dots,2g$ were defined in the previous
section. There are $2g$ independent solutions of (\ref{e11}). In
fact, $d\widetilde{X}$ is necessarily a linear combination of the
basis of harmonic 1-forms plus exact forms. For each $d\widehat{X}^{s}$ there
exists a unique $\phi^{s}$, single-valued over $\Sigma$ such that

\begin{equation} D_{r}D_{r}\widehat{X}^{s}+D_{r}D_{r}\phi^{s}=0. \end{equation}

The most general solution for $\widetilde{X}_{1}$ satisfying
$D_{r}D_{r}\widetilde{X}_{1}=0$ is then

\begin{equation}\label{e12}
d\widetilde{X}_{1}=L_{s}(d\widehat{X}^{s}+d\phi^{s}), \end{equation}

since the only solution in terms of pure exact forms is the trivial
one. We notice also that $d\widehat{X}^{s}+d\phi^{s}$ $s=1,\dots,2g$
are linearly independent. The most general solution for
$\mathcal{D}_{r}\mathcal{D}_{r}\widetilde{X}=0$, at least
perturbatively in $A_{r}$ is of the same form (\ref{e12}) since all
new contributions to the solution are exact. We may rewrite
(\ref{e10}) in
 the form

 \begin{equation}\label{e13}
 dX^{7}=RL_{s}d\widetilde{X}^{s}+d\phi,
 \end{equation}

we notice that $L_{s}$ are the same as in  (\ref{eq1}). The only change is in the exact 1-forms.\\

We may now analyze the contribution of the $dX^{7}$ field to the
hamiltonian. In addition to its conjugate momentum, which appears
quadratically we have

\begin{equation}\label{e14} V_{7}=\left\langle
(\mathcal{D}_{r}X^{7})^{2}+\{X^{m},X^{7}\}^{2}\right\rangle=
\left\langle
(L_{s}\mathcal{D}_{r}\widetilde{X}^{s})^{2}+(\mathcal{D}_{r}\phi)^{2}+
\{X^{m},X^{7}\}^{2}\right\rangle \end{equation}

where we have explicitly used (\ref{e11}).

We then obtain the bound

\begin{equation}\label{e15}
 V_{7}\geq \left\langle
(\mathcal{D}_{r}\phi)^{2}+\{X^{m},X^{7}\}^{2}\right\rangle
\ge (\mathcal{D}_{r}\phi)^{2}
\end{equation}

which directly shows that the winding corresponding to $dX^{7}$ does
not affect the qualitative properties of the spectrum of (\ref{e9}).
The first term in (\ref{e14})

\begin{equation} \left\langle
L_{s}(\mathcal{D}_{r}\widetilde{X}^{s})^{2}\right\rangle \end{equation}

corresponds, for a given $A_{r}$, to the value of the original
quadratic term $\left\langle
(\mathcal{D}_{r}X^{7})^{2}\right\rangle$ in the action evaluated at
a configuration that minimizes it. This is the case, since in the
usual Dirichlet internal product

\begin{equation} (A,B)=\int_{\Sigma}\mathcal{D}_{r}A\mathcal{D}_{r}B
\end{equation}

defined on fields modulo constants, the exact and harmonic functions
are orthogonal. Exactly
 the same decomposition as in (\ref{e14}) occurs when evaluating the sum over isomorphisms class of line bundles $L$ in
the partition function of an abelian gauge field. The sum over $L$
gives a generalized $\theta$ function \cite{witten, caicedo}. The
third term in (23)contributes with $L_{s}^{2}$ to the coefficient of the
quadratic mass terms $(D_{s}X^{m})^{2}$ the total factor
being $(1+L_{s}^{2})$. The first term $\left\langle
(\mathcal{D}_{r}\widetilde{X}^{s})^{2}\right\rangle$ contains a
complicated dependence on $A_{r}$, if one is interested in analyzing
it is better to start with the decomposition (\ref{e10}) instead of (\ref{e13}). For the
purpose of our analysis we may neglect this complicated positive
term and consider the hamiltonian with a potential  contribution given by the
right hand side of (\ref{e15}). The inequality (\ref{e15}) will
ensure that the qualitative properties of the latest are also valid
for the original complete hamiltonian.

We will now construct the dual formulation  to (\ref{e9}) when $dX^{7}$
is restricted by the condition (\ref{e6}) ensuring that $X^{7}$
takes values on $S^{1}$. We follow \cite{caicedo}, given

\begin{equation}\label{16a}
\mathcal{L}=p_{I}\dot{X}^{I}+p\dot{X}-\mathcal{H}(p_{I},X^{I},p,x)
\end{equation}

where

\begin{equation}\label{16b} \oint_{C_{s}}dX=RL_{s}\end{equation}

and the dependence on $X$ is only through its derivatives
$\partial_{\lambda}X$ and construct

\begin{equation}\label{17}
\left\langle\widehat{\mathcal{L}}+W_{\lambda}F_{\mu\nu}\epsilon^{\lambda\mu\nu}\right\rangle
\end{equation}

where

\begin{equation}\label{18}
\widehat{\mathcal{L}}=p_{I}\dot{X}^{I}+pW_{0}-\mathcal{H}(p_{I},X^{I},p,W_{a})
\end{equation}

the elimination of $W_{0}$,  through its field equation or
directly from
 a gaussian integral in the functional integration, yields the dual action

\begin{equation}\label{19a}
\widetilde{\mathcal{L}}=p_{I}\dot{X}^{I}+\Pi^{a}\dot{A}_{a}+A_{0}\partial_{a}\pi^{a}
-\mathcal{H}(p_{I},X^{I},F_{ab}\epsilon^{ab},-\frac{1}{2}\epsilon_{ba}\Pi^{b}).
\end{equation}

It is already in a canonical hamiltonian formulation. The new
hamiltonian is obtained from the original one by making the above
replacements. Notice that there is no assumption on the structure of
$\mathcal{H}$, it is not necessarily quadratic. In our particular case the dependence on $p$ and $W_{a}$ is quadratic. Condition (\ref{16b}) becomes now

\begin{equation}\label{19b}
\oint_{C_{s}}(-\frac{1}{2}\epsilon_{ba}\Pi^{b})d\sigma^{a}=\frac{1}{R}m_{s}
\end{equation}

where $m_{s}$ are integers.

We notice that $A_{s}$ is not a connection in a line bundle over
$\Sigma$. In fact the condition

\begin{equation}\int_{\Sigma}F_{ab}d\sigma^{a}\wedge d\sigma^{b}=2\pi
n
\end{equation}

is not necessarily satisfied. In order to have a connection on line
bundle over $\Sigma$ one should require a periodic euclidean time on
the functional integral formulation. In that case the condition
(\ref{e6}), where now the basis of one-dimensional homology includes
the additional $S^{1}$, ensures that $F_{\mu\nu}$ is the curvature
of a one-form connection over the three dimensional base manifold.
Under this assumption the condition (\ref{e6}) for any $L_{s}$
implies summation over all $U(1)$ principle bundles. The
contribution of this summation of the partition function is a
generalized $\theta$ function \cite{witten} arising from the
evaluation of the abelian action at minimizing configurations, that
is monopole-type solutions \cite{bellorin1}.

 The final expression of the dual formulation to (\ref{e9}) when $X^{7}$ is wrapped on a $S^{1}$, condition (\ref{e6}), is
\begin{equation}\label{e}
\begin{aligned}
H_{d}=&\int_{\Sigma} \sqrt{w}d\sigma^{1}\wedge
d\sigma^{2}[\frac{1}{2}(\frac{P_{m}}{\sqrt{W}})^{2}
+\frac{1}{2}(\frac{\Pi^{r}}{\sqrt{W}})^{2}+\frac{1}{4}\{X^{m},X^{n}\}^{2}+\frac{1}{2}(\mathcal{D}_{r}X^{m})^{2}\\
& \nonumber
+\frac{1}{4}(\mathcal{F}_{rs})^{2}+\frac{1}{2}(F_{ab}\frac{\epsilon^{ab}}{\sqrt{W}})^{2}
+\frac{1}{8}(\frac{\Pi^{c}}{\sqrt{W}}\partial_{c}X^{m})^{2}+\frac{1}{8}[\Pi^{c}\partial_{c}(\widehat{X}_{r}+A_{r})]^{2}]\\
& \nonumber +
\Lambda(\{\frac{P_{m}}{\sqrt{W}},X^{m}\}-\mathcal{D}_{r}(\frac{\Pi^{r}}{\sqrt{W}})
-\frac{\Pi^{c}}{2\sqrt{W}}\partial_{c}(F_{ab}\frac{\epsilon^{ab}}{\sqrt{W}}))+\lambda\partial_{c}\Pi^{c}]\\
\nonumber &+\int_{\Sigma} \sqrt{W}[-\overline{\Psi}\Gamma_{-}\Gamma_{r}\mathcal{D}_{r}\Psi+ \overline \Gamma_{-}\Gamma_{m}\{X^{m},\Psi\}+1/2\overline{\Psi}\Gamma_{7}\Pi^{b}\partial_{b}\Psi]+\Lambda \{\overline{\Psi}\Gamma_{-}, \Psi\}
\end{aligned}
\end{equation}

The term

\begin{equation}
\begin{aligned}
&\frac{1}{2}(\frac{P_{m}}{\sqrt{W}})^{2}+\frac{1}{4}\{X^{m},X^{n}\}^{2}
+\frac{1}{2}(F_{ab}\frac{\epsilon^{ab}}{\sqrt{W}})^{2}
+\frac{1}{8}(\frac{\Pi^{c}}{\sqrt{W}}\partial_{c}X^{m})^{2}\\
& \nonumber +\Lambda(\{\frac{P_{m}}{\sqrt{W}},X^{m}\}
-\frac{1}{2}\Pi^{c}\partial_{c}(F_{ab}\frac{\epsilon^{ab}}{\sqrt{W}}))+\lambda\partial_{c}\Pi^{c}
\end{aligned}
\end{equation}
\\

describe the canonical density of a Dirac-Born Infeld theory in
terms of  $G_{ab}=\partial_{a}X^{m}\partial_{b}X^{m}$
 and $F_{ab}$. In the full theory with the hamiltonian $H_{d}$, there are additional
  interacting terms describing the coupling to the sector wrapped onto a $T^{6}$.
We considered the M2  with all physical configurations wrapped in an irreducible way on a $T^6\times S^1$ target space. If we now take the compactified sector of the target to be $T^7=(S^1)^7$,we should then consider all possible decompositions of the form $T^6\times S^1$.The Hilbert space of physical configurations is then enlarged by considering all possible holomorphic immersions and their corresponding physical states in terms of single valued fields over the base manifold, as explained in section 2. The breaking of susy induced by the ground state follows in the same way.
\section{$N=1$ supersymmetry}
The topological condition associated to the central charge
determines an holomorphic minimal immersion from the $g$-Riemann
surface to the 2g-torus target manifold. This minimal immersion is
directly related to the BPS state that minimizes the hamiltonian.
When we start with the $g=1$ and $T^{2}$ on the target space the
ground state preserves $\frac{1}{2}$ of the original supersymmetry
with parameter a 32-component Majorana spinor. When we consider our
construction for a $g=2,3$ and $T^{4}, T^{6}$ torus on the target,
the analysis of the SUSY preservation becomes \\ exactly the same as
when considering orthogonal intersection of 2-branes with the time
direction as the intersecting direction \cite{smith}. The SUSY of
the ground state preserves $\frac{1}{4}$, $\frac{1}{8}$ of the
original SUSY. The ground state in all these cases corresponds to
\begin{equation} \Psi=0\quad
X^{m}=0\quad X^{r}_{i}=\widehat{X}^{r}_{i} \end{equation}

The preservation of the ground state implies the breaking of the
supersymmetry. In the light cone gauge, we end up when $g=3$ with
$\frac{1}{8}$ of the original SUSY, that is one complex grassmann
parameter corresponding to a $N=1$ light-cone SUSY multiplet.

The action is invariant under the whole light-cone SUSY. There is a
whole class of minima for the hamiltonian, corresponding to
\begin{equation}
\begin{aligned}& \Psi=\epsilon_{1}+\epsilon_{2}\\ \nonumber &
X^{r}=\widehat{X}^{r}+i\overline{\epsilon}_{2}\Gamma\epsilon_{1}\\
\nonumber & X^{m}=i\overline{\epsilon}_{2}\Gamma^{m}\epsilon_{1}.
\end{aligned}
\end{equation}

 However when the vacuum is spontaneously fixed to one of them, the
 SUSY is broken at the quantum level up to $N=1$ when the target is
 $M_{5}\times T^{6}$. There is no further breaking when we
 compactify the additional $S^{1}$, to have a target $M_{4}\times T^{6}\times
 S^{1}$.

\section {Discreteness of the spectrum}
We consider a gauge fixing procedure on a BFV formulation of the
theory. Several gauge conditions are appropriate to analyze the
qualitative properties of the spectrum. We may impose as in
(\cite{bgmmr},\cite{bgmr}) a gauge choice once a normal basis of functions
over $\Sigma$ is introduced in the theory. We
 may otherwise consider a Coulomb gauge condition $D_{r}A_{r}=0$. We may solve it in terms
 of longitudinal and transverse modes in the usual way, together with a resolution of the
first class constraint, the Gauss constraint. In this case once all
the canonical hamiltonian is expressed in terms of the transverse
canonical modes one is left with a positive but complicated term
arising from the square of the momenta terms after the decoupling of
the longitudinal term has been obtained. It is of the form

\begin{equation} D_{r}\Pi^{L}D_{r}\Pi^{L} \end{equation}
 where $\Pi^{L}$ is the longitudinal part of the momenta $\Pi_{r}$, and $\Pi^{L}$
 has to be replaced by the solution of the constraint.
In what follows we may eliminate such positive term, since the
discreteness of the lower bound operator ensures the same property
for the original hamiltonian. The same argument was used in
\cite{bgmmr}.

We may also consider a gauge fixing condition \begin{equation}
\chi\equiv aD_{r}A_{r}+B \end{equation} where $B$ is the BRST
transformed of the antighost field while $a$ is a real number which
can be chosen in a way to cancel the $(D_{r}A_{r})^{2}$ contribution
from the $\mathcal{F}^{2}$ term in the hamiltonian. After a
redefinition of $B$ it decouples from the functional integral, we
end up with a canonical formulation in terms of the
 square of all the momenta together with the quadratic mass terms for each mode in the formulation. An important aspect to mention
 is that in all these cases the ghost fields do not decouple from the action,
however the contributions are always linear on the configuration
variables.
 Theorem 2 in \cite{br} ensure that this ghost contribution does not change the discreteness properties of the canonical formulation.
 The discussion of the spectral properties of the hamiltonian is then largely simplified by those consideration. We may reduce
to the physical degrees of freedom or we may enlarge the phase space
as in BFV
 canonical formalism, in both cases the analysis reduces to a Schr\"oendiger operator
with quadratic mass terms and positive potential. We may consider
for example,using (\ref{e15})

\begin{equation}\begin{aligned} \widehat{H}=&
\int_{\Sigma}\sqrt{W}[\frac{1}{2}(\frac{P_{m}}{\sqrt{W}})^{2}
+\frac{1}{2}(\frac{P}{\sqrt{W}})^{2}+\frac{1}{2}(\frac{\Pi^{r}}{\sqrt{W}})^{2}+\frac{1}{4}\{X^{m},X^{n}\}^{2}\\
&
\nonumber+\frac{1}{4}\{X^{m},X^{7}\}^{2}+\frac{1}{2}(\mathcal{D}_{r}X^{m})^{2}+\frac{1}{2}(\mathcal{D}_{r}\phi)^{2}
+\frac{1}{4}(\mathcal{F}_{rs})^{2}+\frac{1}{2}(\mathcal{D}_{r}A_{r})^{2}]
\end{aligned}
\end{equation}
\\

where $P$ is the conjugate momenta of $\phi$, the contribution of
the compactification on $S^{1}$. This hamiltonian is exactly of the
form considered in \cite{bgmmr},\cite{bgmr}\cite{br}. We may then
apply , for a regularized version of it, the results developed
there.

We consider, in the usual way, a decomposition of all scalar fields
over $\Sigma$ in terms of an orthonormal discrete basis
$Y_{A}(\sigma^{1},\sigma^{2})$. It is relevant in this approach to
consider a compact closed Riemann surface $\Sigma$

\begin{equation}
\begin{aligned}
X^{m}(\Sigma,\sigma, \tau)=&
X^{mA}(\tau)Y_{A}(\sigma^{1},\sigma^{2})\\ \nonumber
A_{r}(\Sigma,\sigma, \tau)=& A^{A}_{r}(\tau)Y_{A}(\sigma^{1},\sigma^{2})\\
\phi(\Sigma,\sigma, \tau)=&
\phi^{A}(\tau)Y_{A}(\sigma^{1},\sigma^{2})
\end{aligned}
\end{equation}
\\

The symplectic bracket is also a scalar over $\Sigma$, hence it must
be rewritten in terms of the basis \begin{equation}
\{Y_{A},Y_{B}\}=f_{AB}^{C}Y_{C}\end{equation}

after defining

\begin{equation}\int_{\Sigma}Y_{A}Y_{B}=\eta_{AB}\end{equation}

we get

\begin{equation} \int_{\Sigma}\{Y_{A},Y_{B}\}Y_{C}=f_{ABC},\end{equation}

$f_{ABC}$ are consequently completely antisymmetric. $f_{AB}^{C}$ are
the structure constant of the area preserving diffeomorphism. We
then replace those expressions into the hamiltonian density and
integrate the $\sigma^{1},\sigma^{2}$ dependence. We obtain then a
formulation of the operator in terms of the $\tau$ dependent modes
only. We now consider a truncation of the operator, that is we
restrict the range of the indices $A,B,C$ to a finite set $N$ and
introduce constants $f_{AB}^{N\quad C}$ such that

\begin{equation} lim_{N\to\infty}f_{AB}^{N\quad C}=f_{AB}^{C} \end{equation}

In \cite{dwln},\cite{dwmn}, $f_{AB}^{N\quad C}$ are the structure
constants of $SU(N)$, that is the truncated theory has also a gauge
symmetry. In \cite{bgmr} for the supermembrane with central charges
compactified on a $T^{2}$ the truncated theory in terms of $SU(N)$
structure constants also has a gauge symmetry. The algebra of first
class constraints in both cases is the same. However, in the proof
of the discretness of the spectrum in \cite{bgmr} the algebraic
properties of $f_{AB}^{N\quad C}$ do not play any role at all. We
then proceed to the analysis of the spectrum of the truncated
Schr\"oedinger
operator associated to $\widehat{H}$ without further requirements on the constants $f^{N}$:\\
 i) The potential of the Schr\"oedinger  operator only vanishes at
the origin of the configuration space:

\begin{equation} V=0 \to \vert\vert (X,A,\phi)\vert\vert=0 \end{equation}

where $\vert\vert .\vert\vert$ denotes the euclidean norm in
$R^{L}$. We notice that the original hamiltonian as well as
$\widehat{H}$ are defined on fields up to constants.

ii) There exists a constant $M> 0$ such that

\begin{equation} V(X,A,\phi)\ge M\vert\vert (X,A,\phi))\vert\vert^{2}\end{equation}

Again, this bound arises from very general considerations. In fact,
writing $(X,A,\phi)$ in polar coordinates

\begin{equation}
X=Rx\quad A=Ra\quad \phi=R\varphi \end{equation}
 where $\theta\equiv (x,a,\varphi)$ is defined on the unit sphere, $\vert\vert(x,a,\varphi)\vert\vert=1$,
we obtain

\begin{equation} V(X,A,\phi)=R^{2}P_{\theta}(R) \end{equation}

where \begin{equation} P_{\theta}(R)=R^{2}k_{1}(\theta)+R
k_{2}(\theta)+k_{3}(\theta)>0 \end{equation}

with $k_{3}(\theta)> 0$, $k_{1}(\theta)\ge 0$ and
$k_{1}(\theta)=0\Rightarrow k_{2}(\theta)=0$. We then define
\begin{equation} \mu(\theta)=min_{R}P_{\theta}(R), \end{equation}

it is  continuous in $\theta$ and $\mu(\theta)>0$. Using the
compactness of the unit sphere we obtain

\begin{equation} V(X,A,\phi)=R^{2}P_{\theta}(R^{2})\ge R^{2}min_{\theta}=M R^{2}
\end{equation}

The Schr\"oedinger operator is then bounded by an harmonic
oscillator. Consequently
it has a compact resolvent. We now use theorem 2 \cite{br} to show that \\

i) The ghost and antighost  contributions to the effective action,\\

ii) the fermionic contribution to the susy hamiltonian,\\

do not change the qualitative properties of the spectrum of the
hamiltonian. In fact, both contributions are linear on the
configuration variables.

In addition the susy contribution cancels the zero point energy of
the bosonic oscillators even in the exact theory
\cite{stelle}, \cite{bgmr2}. \\

We have shown then that the regularized compactified on the target
space $M_{4}\times T^{6}\times S^{1}$ has a compact resolvent and
hence a discrete spectrum with finite multiplicity. We expect the
same result to be valid for the exact theory.

\section{Physical properties}
So far we have seeing that the action of the $N=1$ supermembrane
 in four dimensions has a regularized discrete
spectrum.

One of the characteristics of the theory is that due to the
topological condition the fields acquire mass. This facts represents
an alternative to Higgs mechanism since no Higgs particle is
involved. There has been several mass generating mechanisms
\cite{scherk}-\cite{shaposhnikov} in the literature. Since ours does
not correspond strictly speaking to non of them although there are
some resemblances, we will explain briefly just for the sake of
clarity.
 In here, the fields of the theory $X^{m},A_{r},\phi$ acquire mass via
the vector fields $\widehat{X}_{r}$ defined on the supermembrane.
Since those fields do not live in the target-space there is no
violation of Lorentz invariance. In fact in resemblance with
Scherk-Schwarz mechanism
 they induce a monodromy on the fields.  It is important to point out that the
number of degrees of freedom in 11D and in 4D is preserved, but just
redistributed, contrary to the case of an effective compactification
a la KK,  where a tower of fields appear. This fact has the a
advantage for many phenomenological purposes of maintaining the
number of fields
small.\\
\newline

 One possible extra question is the analysis of moduli stabilization.
Moduli are massless scalar fields that may parametrize the
compactified geometry as well as different matter sectors. It can be
distinguished two types of moduli: quantum moduli and classical
moduli.\\

 Quantum moduli of the theory, generically is not known
although there has been some approximations for particular set-ups
in which the different bunch of these moduli show the interpolation
between different vacua with different gauge groups \cite{friedman}.
As we have pointed out along the text, the theory of quantum
supermembranes with central charges we have constructed depends then
on the moduli space of compact Riemanian surfaces $M_{g}$ only. It
is defined on the conformal equivalent classes of compact Riemann
surfaces, and also depends on the moduli identifying the holomorphic
immersion from $M_{g}$ to the target manifold.

Generically the analysis of moduli fields have been performed at
classical level in a supergravity approach
\cite{acharya}-\cite{diana}. It has being performed in effective 4D
potentials of the supergravity approximations of M-inspired actions.
The K\"ahler potential is expressed in terms of them
\cite{lukas},\cite{dall'agata}. Since our approach is exact these
terms do not appear, however the action posseses  scalars that may
lead to flat directions in the potential. We can distinguish between
two types of scalars fields, those associated to the position of the
supermembrane in the transverse dimensions, - analogous to what in
String theory represent the open string moduli- and the scalars
 whose vevs parametrize the compact manifold - analogous to what in
String theory represent the closed string moduli-.\\

We are going to analyze separately the two types of classical moduli. This
decoupling approach is only justified iff the scales of
stabilization (the masses of the moduli) are clearly different,
otherwise the minimization with respect to the whole set of moduli
(geometrical and of matter origin) should be performed. An
exhaustive analysis of this fact is beyond the scope of this
article.
However some considerations still can be made:\\

At classical level the behaviour of the theory is known. The theory
does not contain any string-like configuration. Let assume for the
moment a compact manifold whose radii $R_{1},\dots,R_{7}$ are fixed.
The $X^{m}$ that parametrize the position in the transverse
dimensions of the supermembrane acquire mass due to the central
charge condition so there are no flat directions in the scalar piece
of the potential. The 7-component component has an  induced effect due to
the central charge condition through the quadratic coupling with the simplectic gauge fields
$A_{r}$ and gain also an
effective mass. All of these type of moduli becomes stable.\\

So far we have considered the 7-torus to be rigid, in such a way
that the $R_{1}\dots,R_{7}$ are kept fixed and they do not appear in
the metric (they would be the putative closed string moduli). If now
we relax this condition and let them  vary smoothly, we can ask
ourselves if in principle it is possible to obtain a minimum. An
heuristical argument to support moduli stabilization is the
following: We are dealing in our construction with non trivial gauge
bundles that can be represented as worldvolume fluxes
$\cite{d2-d0}$. Since for construction the mapping represent a
minimal immersion on the target space they induce a similar effect
that the one induced by the generalized calibration. Minimal
calibrations take also into account the dependence on the base
manifold, the Rieman surface chosen $\Sigma$. The condition of the
generalized calibration -which shows the deformation of the cycles
that are wrapped by the supermembrane- represent a condition for
minimizing the energy \cite{evslin}. It happens the same with the
minimal inmersions. For a given induced flux,one may expect the
volume to be fixed. The supermembrane with central charges is
wrapping all of the 7-torus with the maximal amount of monopoles
induced on it, so the overall geometric moduli in principle will be
also stabilized.
\newline

Let us illustrate it, with the particular case of an isotropic
torus, i.e. $R_{1}=\dots=R_{7}=R_{0}$. We then obtain for the
potential in $\widehat{H}$, the following expression

\begin{equation} V=A+ BR_{0}+CR^{2}_{0}+DR^{4}_{0} \end{equation}

where $A\ge 0$, $C\ge 0$ and $D>0$, with the following expressions
\begin{equation}
\begin{aligned}
&D=\frac{1}{4}\int_{\Sigma}\sqrt{W}\{\widehat{X}^{r},\widehat{X}^{s}\}^{2}\\
\nonumber & C=\frac{1}{2}\int_{\Sigma}\sqrt{W}[(D_{r}X^{m})^{2}+(D_{r}\phi)^{2}+(D_{r}A_{s})^{2}+\{X^{m},\widehat{X}^{s}L_{s}\}^{2}]\\
\nonumber & B=\int_{\Sigma} \sqrt{W}[\frac{1}{2}\{X^{m},L_{s}\widetilde{X}^{s}\}\{X^{m},\phi\}+
D_{r}X^{m}\{A_{r},X^{m}\}+D_{r}\phi\{A_{r},\phi\}+ \frac{1}{2}(D_{r}A_{s}-D_{s}A_{r})\{A_{r},A_{s}\}]\\
\nonumber &
A=\int_{\Sigma}\sqrt{W}[\frac{1}{4}\{X^{m},X^{n}\}^{2}+\frac{1}{4}\{X^{m},\phi\}^{2}
+\frac{1}{2}\{A_{r},X^{m}\}^{2}+\frac{1}{2}\{A_{r},\phi\}^{2}+\{A_{r},A_{s}\}^{2}]
\end{aligned}
\end{equation}
where we have extracted the $R_{0}$ factor from the expression of
the derivative $D_{r}$.

 The quadratic mass terms contribute to the expression of $C$. $C$
 is zero if and only if $(X^{m},\phi,A_{s})$ are constants, in
 the equivalence class of zero.
We obtain

\begin{equation} \frac{d^{2}V}{dR^{2}_{0}}=2C+12DR^{2}_{0}>0 \end{equation}

consequently the problem is always stable with respect to the
variations of $R$. We may have two possible minima: a minimum
centered at $R_{0}=0$, or a minimum for $R_{0}\ne 0$. The potential
is
globally stable with respect to the modulus $R_{0}$.\\

In the cases in which the radii are all different a more exhaustive
analysis is need. We will consider it elsewhere.
\section{Conclusion}

We obtained the action of the D=11 supermembrane compactified on
$T^6 \times S^1$ with nontrivial central charge induced by a topological
condition  invariant under  supersymmetric and kappa symmetry
transformations. The  hamiltonian in the LCG is invariant under
conformal  transformations on the Riemann surface base manifold. The
susy is spontaneously broken, by the vacuum to  $1/8$ of the
original one. It corresponds in 4D to a $N=1$ multiplet. Classicaly
the hamiltonian does not contain singular configurations and at the
quantum level the regularized hamiltonian has a discrete spectrum,
with finite multiplicity. Its resolvent is compact. The potential
does not contain any flat direction on configuration space nor on
the moduli space of parameters. The hamiltonian is stable on both
spaces. It is stable as a Schrodinger operator on configuration
space and it is structurable stable  on the moduli space of
parameters.

\section{Acknowledgements}
We would like to thank  for helpful conversations to G.~Dall'Agata,
M.~Billo, M.~Frau, J.~Ovalle and H.~Nicolai. M.P.G.M. is partially
supported by the European Comunity's Human Potential Programme under
contract MRTN-CT-2004-005104 and by the Italian MUR under contracts
PRIN-2005023102 and PRIN-2005024045. J.M.~Pena research is carried
by the Ph.D grant 'programa Mision Ciencia', Caracas, Venezuela. The
work of A. R. is partially supported by a grant from MPG, Albert
Einstein Institute, Germany and by PROSUL, under contrat CNPq
490134/2006-08.


\begin{thebibliography}{99}

\bibitem{dwhn} B. de Wit, J. Hoppe and H. Nicolai, {\em On the quantum mechanics of
supermembranes}, Nucl. Phys. {\bf B305} (1988) 545.

\bibitem{dwln} B. de Wit, M. Luscher and H. Nicolai, {\em The supermembrane is unstable}, Nucl. Phys. {\bf B320} (1989) 135.

\bibitem{dwmn} B. de Wit, U. Marquard and H. Nicolai,
{\em Area preserving diffeomorphisms and supermembrane lorentz
invariance}, Commun. Math. Phys. {\bf 128} (1990) 39-62.


\bibitem{dwpp} B. de Wit, K. Peeters and J. Plefka,
{\em Supermembranes with winding.} Phys. Lett.{\bf B409} (1997) 117-123,
 [arXiv: hep-th/9705225].


\bibitem{gmr} M.P. Garcia del Moral and A. Restuccia, {\em On the
spectrum of a noncommutative formulation of the D=11 supermembrane
with winding}, Phys.Rev. {\bf D66}  (2002) 045023,[arXiv: hep-th/0103261].


\bibitem{bgmmr} L. Boulton, M. P. Garcia del Moral, I.
Martin and A. Restuccia,  {\em On the spectrum of a matrix model for the
D=11 supermembrane compactified on a torus with non-trivial
winding}, Class. Quant. Grav. {\bf 19} (2002) 2951,[arXiv:hep-th/0109153].


\bibitem{torrealba} I. Martin, A. Restuccia and R. S. Torrealba,
{\em On the stability of compactified D = 11 supermembranes},
 Nucl. Phys. {\bf B521} (1998) 117-128,[arXiv:hep-th/9706090].

\bibitem{bgmr} L. Boulton, M.P.
Garcia del Moral and A. Restuccia, {\em Discreteness of the spectrum of
the compactified D=11 supermembrane with non-trivial winding},
 Nucl.Phys. {\bf B671} (2003) 343-358,[arXiv:hep-th/0211047].

\bibitem{bellorin} J. Bellorin and A. Restuccia,
{\em D=11 Supermembrane wrapped on calibrated submanifolds},
Nucl.Phys. {\bf B737} (2006) 190-208,[arXiv:hep-th/0510259].


\bibitem{bgmr2}L. Boulton, M.P. Garcia del Moral and A. Restuccia,
{\em The Supermembrane with central charges: (2+1)-D NCSYM,
confinement and phase transition}, Nucl.Phys. {\bf B795} (2008) 27-51,2008
[arXiv:hep-th/0609054].


\bibitem {gutowski} J. Gutowski and G. Papadopoulos,
{\em Moduli spaces and brane solitons for M theory compactifications
on holonomy G(2) manifolds}, Nucl.Phys. {\bf B615} (2001) 237-265,
[arXiv:hep-th/0104105].

\bibitem {lukas} A. Lukas and S. Morris, {\em Moduli Kahler potential for M theory on a G(2) manifold}, Phys.Rev. {\bf D69} (2004) 066003,[arXiv:hep-th/0305078].

\bibitem{acharya} B.S. Acharya, {\em A Moduli fixing mechanism in M theory},
[arXiv:hep-th/0212294].

\bibitem{dall'agata} G. Dall'Agata and N. Prezas, {Scherk-Schwarz reduction of M-theory on $G2$-manifolds with fluxes},
JHEP {\bf 0510} (2005) 103,[arXiv:hep-th/0509052].


\bibitem{diana}B. Acharya, K. Bobkov, G. Kane, P. Kumar and D. Vaman,
 {\em  An M theory Solution to the Hierarchy Problem},
 Phys.Rev.Lett. {\bf 97} (2006) 191601,[arXiv: hep-th/0606262];\\
 B. Acharya, K. Bobkov, G. Kane, P. Kumar and J. Shao,{\em Explaining the Electroweak Scale and Stabilizing Moduli in M
Theory}, [arXiv:hep-th/0701034].

\bibitem{ovalle}I. Martin, J. Ovalle and A. Restuccia, {\em D-branes, symplectomorphisms and noncommutative gauge theories.}
Nucl. Phys. Proc. Suppl. {\bf 102} (2001) 169-175, {\em Compactified
D = 11 supermembranes and symplectic noncommutative gauge theories
}, Phys. Rev. {\bf D64} (2001) 046001, [arXiv:hep-th/0101236].

\bibitem{belhaj} A. Belhaj, M.P. Garcia del Moral, I. Martin, A.
Segui, J. Veiro and A. Restuccia, {\em The Supermembrane with Central Charges on a G2 Manifold}, [arXiv:hep-th/0803.1827].

\bibitem{bst} E. Bergshoeff, E. Sezgin and  P.K. Townsend,
{\em Supermembranes and eleven-dimensional supergravity}, Phys.
Lett. {\bf B189} (1987) 75-78.

\bibitem{witten} E. Witten {\em  On S duality in Abelian gauge theory},
 Selecta Math.{\bf 1} (1995) 383,[arXiv:hep-th/9505186].

\bibitem{caicedo} M.I. Caicedo, I. Martin and A. Restuccia,
{\em  Gerbes and duality.} Annals Phys. {\bf 300} (2002) 32-53,
[arXiv:hep-th/0205002].

\bibitem{bellorin1} J. Bellorin, A. Restuccia
 {\em Extended selfdual configurations as stable exact solutions in Born-Infeld theory},
 Phys.Rev. {\bf D64} (2001) 106003,[arXiv:hep-th/0007066].

\bibitem{smith} D. J. Smith {\em Intersecting brane solutions in string
and M theory.} Class.Quant.Grav. {\bf 20} (2003) R233,[arXiv:hep-th/0210157].

\bibitem{br} L. Boulton and A. Restuccia,
{\em The Heat kernel of the compactified D=11 supermembrane with
non-trivial winding}, Nucl. Phys. {\bf B724} (2005) 380-396,
[arXiv:hep-th/0405216].


\bibitem{stelle} M.J. Duff, T.
Inami, C.N. Pope, E. Sezgin and  K.S. Stelle, {\em Semiclassical
Quantization Of The Supermembrane}, Nucl. Phys. {\bf B297} (1988) 515,


\bibitem{scherk}J. Scherk and J.H. Schwarz {\em How to get masses from extra dimensions}, Nucl.Phys.{\bf B153} (1979) 61-68.

\bibitem{hosotani} Y. Hosotani {\em Dynamical mass generation by compact extra
dimensions}, Phys.Lett{\bf B126} (1983) 309.

\bibitem{sakamoto} M.Sakamoto, M. Tachibana, K. Takegana {\em Spontaneous supersymmetry breaking
from extra dimensions}, Phys.Lett. {\bf B458} (1999) 231-236,[arXiv:hep-th/9902070].

\bibitem{shaposhnikov} M.E. Shaposhnikov and  P. Tinyakov {\em Extra
dimensions as an alternative to Higgs mechanism}, Phys.Lett. {\bf
B515} (2001) 442-446,[arXiv:hep-th/0102161].


\bibitem{friedman} T. Friedmann {\em On the quantum moduli space of M theory compactifications},
Nucl.Phys.{\bf B635} (2002) 384-394,[arXiv:hep-th/0203256].


\bibitem{d2-d0} M.P. Garcia del Moral, A. Restuccia,
{\em The Supermembrane with central charge as a bundle of D2 - D0
branes}, Institute of Physics Conference Series 2005, Vol {\bf 43},
151, [arXiv:hep-th/0410288].

\bibitem{evslin} J. Evslin and L. Martucci {\em D-brane networks in flux vacua, generalized cycles and calibrations},
JHEP {\bf 0707} (2007) 040,[arXiv:hep-th/0703129].

\end{thebibliography}
\end{document}